# Observation of coherently modulated thermal phonon band and lifetime in superlattice

Yuxuan Liao[1], Hiroshi Uchiyama[2], Naomi Nagai[3], Natália Morais[3,6], Taiushun Manjo[2], Rulei Guo[1], Harsh Chandra[1], Ryohei Nagahiro[1], Bin Xu[1,5], Hiroshi Fukui[2], Daisuke Ishikawa[2], Alfred Q.R. Baron[2,4], Yasuhiko Arakawa[3,6], Kazuhiko Hirakawa[3,6], and Junichiro Shiomi[1,5]*

[1]Department of Mechanical Engineering, The University of Tokyo; Bunkyo-ku, Tokyo, 113-8656, Japan.

[2]Japan Synchrotron Radiation Research Institute (JASRI), SPring-8; 1-1-1 Koto, Sayo, Hyogo 679-5198, Japan.

[3]Institute of Industrial Science, University of Tokyo; Meguro-ku, Tokyo 153-8505, Japan.

[4]Materials Dynamics Laboratory, RIKEN SPring-8 Center, SPring-8; 1-1-1 Koto, Sayo, Hyogo 679-5148, Japan.

[5]Institute of Engineering Innovation, The University of Tokyo; Bunkyo-ku, Tokyo, 113-8656, Japan.

[6]Institute for Nano Quantum Information Electronics, The University of Tokyo, Meguro-ku, Tokyo 153-8505, Japan.

*Corresponding author. Email: Junichiro Shiomi (shiomi@photon.t.u-tokyo.ac.jp);

**Abstract:** Manipulating phonon band structures through wave interference in artificial periodic nanostructures is of fundamental interest and forms the foundation of phonon band engineering. Yet, direct observation of thermal phonon wave interference has remained elusive despite half a century of extensive research. Here, utilizing inelastic x-ray scattering, we directly observed momentum-resolved coherently modulated thermal phonon bands and lifetimes in a short-period GaAs/AlAs superlattice at 300 K. The modulation persists at even 500 K, demonstrating robust coherent thermal phonon transport at high temperatures. Our measurements, combined with *ab initio* lattice-dynamics calculations, further show that thermal phonon wave interference expands the three-phonon scattering phase space and strengthens anharmonic effects, highlighting the interplay between phonon band modulation and anharmonicity. Our observations establish nanostructuring as a versatile route for engineering thermal phonon bands and lifetimes at high temperatures across diverse applications.

**One-Sentence Summary**: Direct observation of coherent modulation of thermal phonon bands and lifetimes at high temperatures

**Main Text:** Phonon band structures in crystals govern important physical phenomena, such as phonon-phonon interactions[1-3] and electron-phonon coupling[4,5], making it crucial in many modern technological revolutions, from superconductivity[6-8], atomic-scale phase transition[9-11], to solid-state thermal physics[2,3,12,13]. Phonon band structures are typically determined by the intrinsic lattice structures. In artificial periodic nanostructures, however, specular phonon scattering at smooth interfaces and surfaces can give rise to phonon wave interference, thereby coherently modulating phonon bands[14,15]. Analogous to the behavior of photons in photonic crystals[16-18], electrons in semiconductors[19-21], and sound waves in phononic crystals[14,22-26], this scientifically significant phenomenon facilitates the artificial engineering of phonon bands through nanostructuring, thereby establishing fundamentally new approaches to manipulate thermal energy transport.

Despite its significance, direct observation of coherently modulated thermal phonon bands and lifetimes remains challenging. Early experimental studies in the 1980s utilizing Raman spectroscopy[27,28], superconducting tunnel junctions[29,30], and laser pump-probe techniques[29,31] have detected energy shifts associated with coherent phonons in superlattices. These measurements, however, only probe phonons near the Γ point, which contribute negligibly to heat conduction. Moreover, the Γ-point phonons probed by these techniques can have wavelengths of hundreds of nanometers—orders of magnitude longer than the few-nanometer wavelengths of thermal phonons at 300 K—and are therefore much more likely to retain coherence. The coherence observed near the Γ point cannot be directly extrapolated to the thermal phonons that carry heat. Whether thermal phonons exhibit wave interference at room temperature remains an open question.

Over recent two decades, subsequent studies have sought evidence of thermal phonon wave interference from variations in lattice thermal conductivity ($\kappa$). Coherently modulated thermal phonon bands reduce group velocity ($v_g$) and thereby suppress $\kappa$ ($\propto v_g^2$), motivating extensive measurements in engineered nanostructures. These measurements include reports of a linearly increased $\kappa$ with increasing superlattice thickness[32], a minimum in $\kappa$ as a function of interface density [33], and lower $\kappa$ in ordered phononic structures compared to disordered ones[34]. In these studies, however, coherent phonon band modulation was inferred indirectly from $\kappa$ variations, primarily at low temperatures. Direct evidence for the persistence of thermal phonon wave interference at and above room temperature has remained elusive, let alone momentum-resolved measurements of coherently modulated thermal phonons bands and lifetimes.

In this study, using high-resolution inelastic X-ray scattering (IXS), we directly observed coherently modulated thermal phonon band structure with phononic band gaps in short-period GaAs/AlAs superlattices at 300 K and 500 K. Our observation provides the first direct evidence of thermal phonon coherence at and above room temperatures, marking a substantial advancement towards artificial engineering of phonon band structures. Combining our experimental observations with *ab initio* lattice dynamics, we further found that coherent phonon band modulation strengthens anharmonic effects such as three-phonon scattering and optical phonon softening, revealing an interplay between phonon band modulation and anharmonicity. Our observations establish nanostructuring as a versatile route for engineering thermal phonon bands and anharmonic phonon-phonon scattering at high temperatures, with extensive implications for phononic metamaterials, microelectronics, and thermoelectrics.

## Superlattice fabrication and characterization

The GaAs/AlAs superlattice has emerged as the most promising system for facilitating phonon coherence[2]. We grew high-quality GaAs/AlAs superlattices comprising 1328 periods coherently on a (0 0 1)-oriented, non-doped single-crystal GaAs substrate using molecular-beam epitaxy (MBE) (**Fig. 1a**). The periodicity was minimized to 2.26 nm, comprising two unit-cell layers of GaAs (1.13 nm) and AlAs (1.13 nm), to obtain a sufficient coherent system. The high-angle annular dark-field scanning transmission electron microscopy (HAADF-STEM) image along (1 1 0)-direction confirmed the high quality of the superlattice (**Fig. 1b**). The *Z*-contrast image, where the measured signal intensity highly correlates with the square of atomic number, allows for identifying atom types and positions. The brighter spots in **Fig. 1b** correspond to Ga or As atoms, whereas the relatively darker ones indicate Al atoms. From the periodic arrangement of the darker spots, we determined the superlattice periodicity to be four cubic lattice unit cells for GaAs or AlAs (2.26 nm). Furthermore, minimal interdiffusion between Ga and Al atoms near the interface, observed within several-nanometer regions, suggests atomically smooth interfaces in the fabricated superlattice.

To further characterize the crystallographic structure of the superlattice, we measured its structure factor $S(\mathbf{Q})$ via X-ray Bragg diffraction along the (0 0 1) direction, where **Q** represents the lattice vector in the reciprocal unit (r.l.u.). The X-ray incidence angle ($\alpha$ in **Fig. 1a**) was optimized to 0.3° to mitigate X-ray penetration into the substrate and reduce background noise[35] (see Methods and **Fig. S1**). **Figure 1c** shows the measured $S(\mathbf{Q})$, where the intensity is proportional to the square of the spatial Fourier transformation of atom coordinates **R** (i.e., $|e^{-i\mathbf{Q}\cdot\mathbf{R}}|^2$). **Q** is mapped to the reduced lattice vector ($\mathbf{q}$=(0 0 $q$)) in the first Brillouin zone (BZ) with the reciprocal lattice vector of **G** (i.e., $\mathbf{Q} = \mathbf{q} + \mathbf{G}$). The Bragg peaks in $S(\mathbf{Q})$ ($q$ = 0.25, 0.49, 0.73, and 0.97) correspond to X-ray path lengths that are integer ($N$) multiples of the lattice constant ($a$) in real space ($Na$), or the reciprocal lattice points that define superlattice BZs in reciprocal space. Consequently, the average superlattice periodicity was determined to be four cubic unit cells of GaAs or AlAs (2.26 nm). The measured Bragg peak positions deviate only slightly from those of the perfect reference superlattice ($q$ = 0.25, 0.50, 0.75, and 1.00), indicating a homogeneous superlattice structure throughout the sample and negligible strain effects.

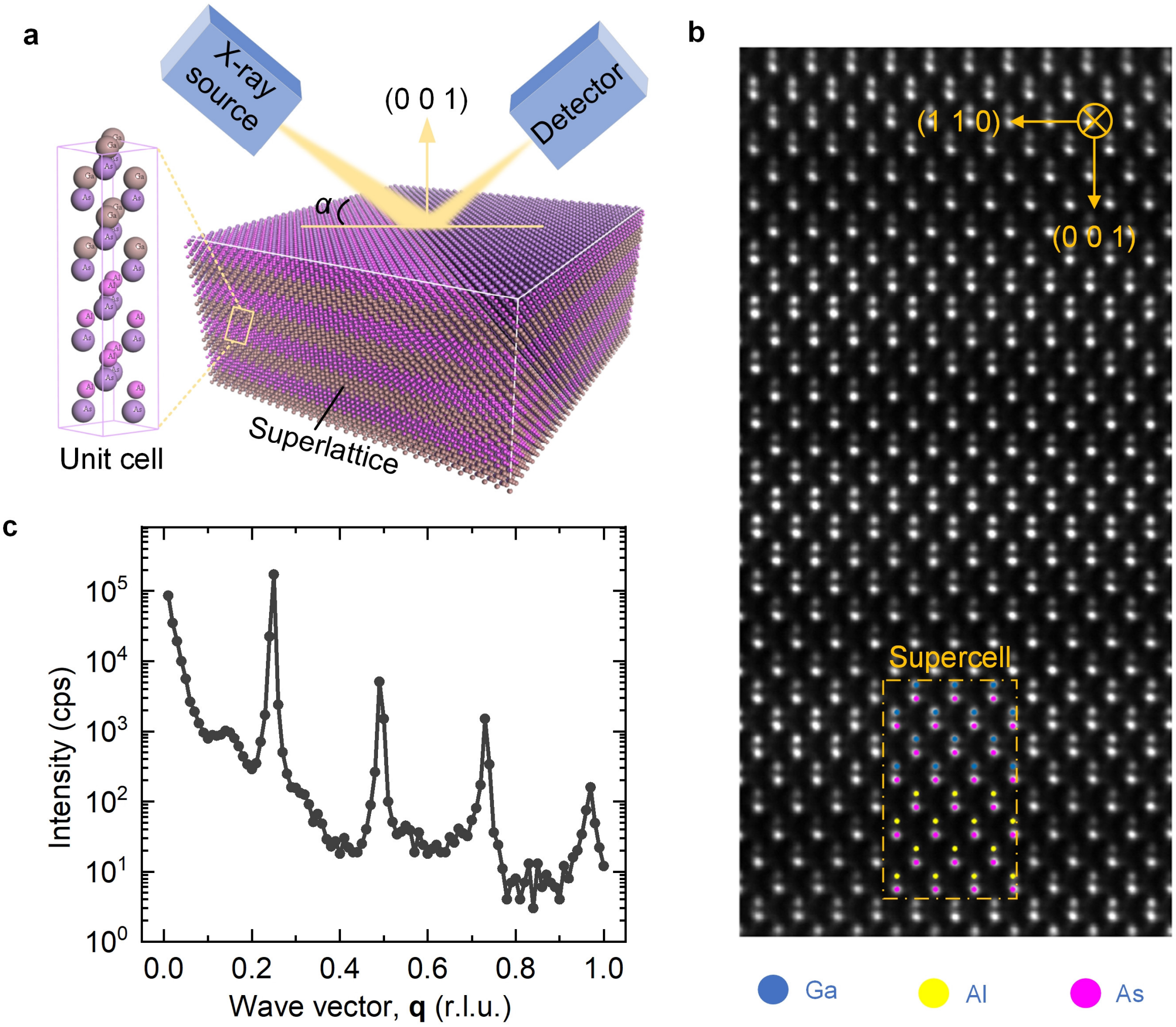

**Fig. 1. X-ray scattering experimental setup and GaAs/AlAs superlattice structure characterization**. (**a**) Schematic of inelastic X-ray scattering (IXS) setup and the superlattice unit cell. Here, $\alpha$ is the X-ray incidence angle with respect to the superlattice surface. The superlattice unit cell, comprising two unit-cell layers of GaAs (1.13 nm) and AlAs (1.13 nm), has 32 atoms. (**b**) High-resolution annular dark field scanning transmission electron microscopy (ADF-STEM) image of the superlattice, where a supercell of 2 × 2 × 1 of the superlattice unit cell is marked. The scale bar reflects the interatomic distance of As atoms in (0 0 1) direction (measured at 5.654 Å). (**c**) X-ray diffraction intensity of the superlattice for $\mathbf{Q} = (1\ 1\ 5\text{-}q)$, with $\mathbf{q} = (0\ 0\ q)$.

The HAADF-STEM and X-ray Bragg diffraction characterizations demonstrate the high quality of the superlattices. The atomically smooth interfaces scatter phonons specularly with preserved phases, whereas the short periodicity eliminates anharmonic dephasing inside each GaAs and AlAs layer, making the superlattice an ideal coherent structure for phonon wave interference.

## Observation of coherently modulated phonon band structure

Our high-resolution IXS experiment utilized high-energy (21.747 keV) photons, enabling the Ewald sphere to encompass many superlattice BZs. The IXS signal intensity exhibited no periodic variation across BZs. Therefore, we employed density functional perturbation theory (DFPT) to rigorously calculate the coherently modulated phonon band structure and the corresponding dynamic structure factor $S(\mathbf{Q},\omega)$ across all measurable BZs, where $\omega$ represents the energy transfer or phonon energy. We focused on the high-symmetry Γ-X (0 0 1) direction, which exhibits a simpler phonon band structure due to the degeneracy of the two transverse branches.

**Figure S2** illustrates a close agreement between calculated and experimental phonon band structures for bulk GaAs and AlAs along high-symmetry lines in the first BZ, confirming the accuracy of DFPT in predicting the phonon band structure of Ga-Al-As systems. In contrast to bulk GaAs and AlAs, the coherent superlattice structure flattens optical phonon bands across the entire BZ (**Fig. 2a**). Consequently, the total phonon density of states deviates from the average of those of bulk GaAs and AlAs. While bulk GaAs and AlAs have similar acoustic phonon branches due to their analogous crystal structures, the superlattice exhibits phononic band gaps of approximately 1 meV near the edges of its BZs ($q$ = 0.125 plus integer multiples of 0.25; marked by red arrows in **Fig. 2a**). The superlattice longitudinal branches are intricately modulated by these band gaps, and lie between those of bulk GaAs and AlAs. Notably, the superlattice transverse branches are flattened near 10 meV, where there is a mismatch between the phonon energies of bulk GaAs and AlAs.

The phonon $S(\mathbf{Q},\omega)$ determines the signal intensity in IXS experiments. Its **Q**-dependence allows transverse and longitudinal branches to be selectively measured in specific BZs. **Figures 2b-c and 2d-e** show the calculated $S(\mathbf{Q},\omega)$ in two selected BZs that enhance longitudinal and transverse modes, respectively. The colormap intensity is proportional to $S(\mathbf{Q},\omega)$ values. As the phonon eigenmodes are projected onto **Q** and the phase factor $|e^{-i\mathbf{Q}\cdot\mathbf{R}}|^2$ appears in $S(\mathbf{Q},\omega)$, most phonon modes have negligible $S(\mathbf{Q},\omega)$. Only the phonon modes visible in **Figs. 2b-e** have sufficient signal intensity to be measured by IXS. Consequently, we target these 'visible' coherently modulated phonon bands, distinct from those of the two bulk materials.

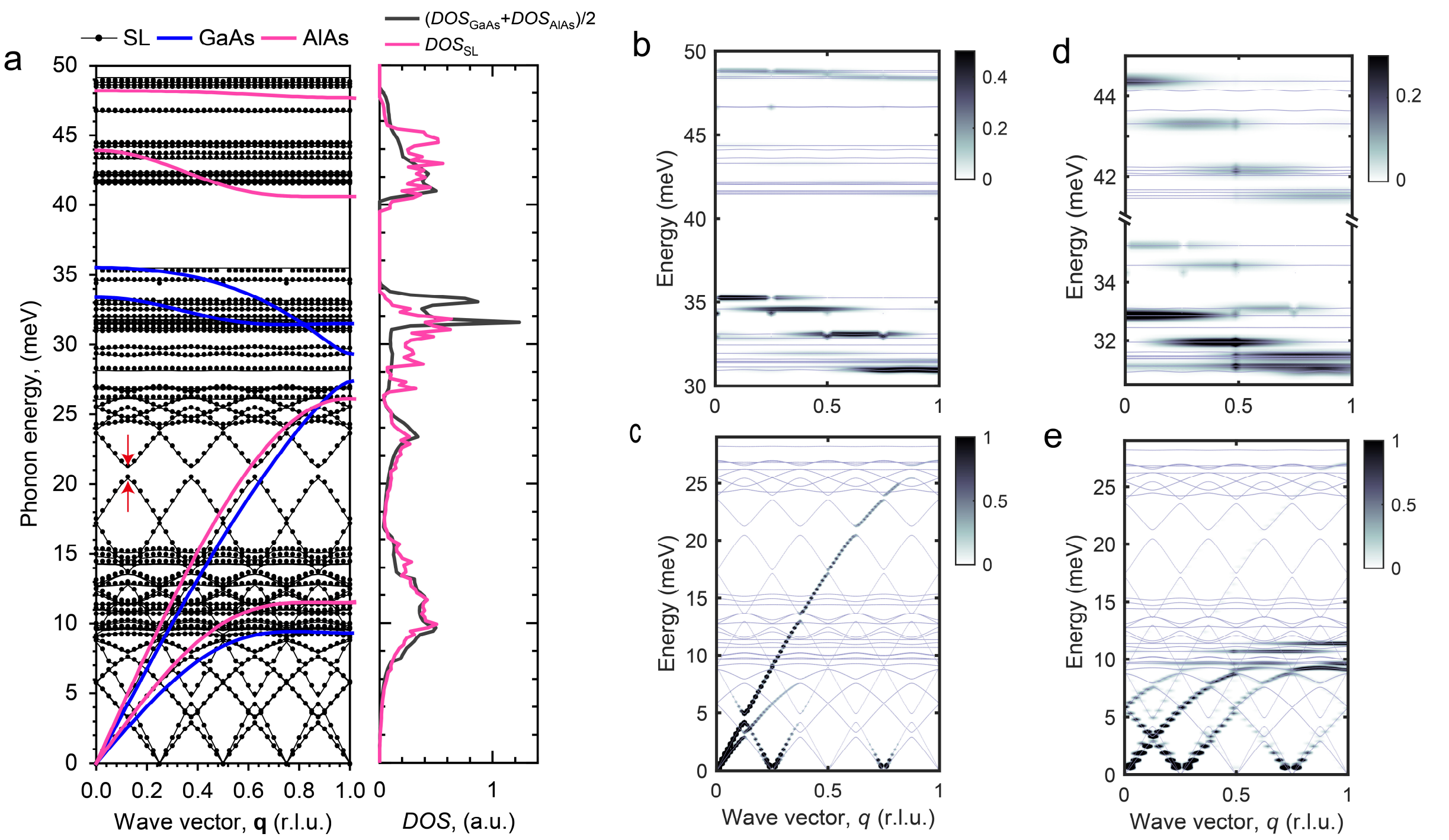

**Fig. 2. Calculated phonon band structure and dynamic structure factor along Γ-X at 0 K. (a)** Phonon band structures of GaAs and AlAs, and the superlattice (SL). The red arrows show an example of a band gap at $q$ = 0.125. The total phonon density of states ($DOS$) of bulk GaAs ($DOS_{GaAs}$) and AlAs ($DOS_{AlAs}$) are averaged for comparison with that of the superlattice ($DOS_{SL}$). **(b-c)** Phonon band structure (gray lines) and the longitudinal-mode-enhanced phonon $S(\mathbf{Q},\omega)$ (contour plot) for $\mathbf{Q}$ = (1 1 5–$q$) in the superlattice. **(d-e)** Phonon band structure (gray lines) and the transversal-mode-enhanced phonon $S(\mathbf{Q},\omega)$ (contour plot) for $\mathbf{Q}$ = (4 4 2+$q$) in the superlattice. The value of calculated $S(\mathbf{Q},\omega)$ in **Fig. 2b-e** is proportional to colormap values.

The high-resolution IXS experiments were conducted at the BL35XU beamline of the SPring-8 synchrotron facility[36-38]. The spectrometer was configured with an energy resolution of 1.4–1.5 meV (full-width at half-maximum) and a cross-plane momentum window of 0.03 in reciprocal lattice units. The X-ray incidence angle was maintained at 0.3°. We measured longitudinal phonon branches at $\mathbf{Q}$ = (1 1 5-$q$), where $S(\mathbf{Q},\omega)$ for transverse modes was negligible (**Fig. 2b-c**). Similarly, we measured transverse branches at $\mathbf{Q}$ = (4 2 2+$q$) and (4 4 2+$q$) (**Fig. 2d-e**). The energy and linewidth ($\sigma$) of each measured phonon were obtained by fitting its IXS energy spectrum to Eq (1) in the Method Section in the Supplementary Materials[35]. The IXS experiments were conducted at 300 and 500 K.

**Figure 3a** shows the transverse optical phonon bands measured at 300 K for $\mathbf{Q}$ = (4 2 2+$q$) and (4 4 2+$q$), represented by filled red circles and square dots, respectively. Consistent with DFPT predictions, the measured phonon bands exhibit flattening across the entire $\mathbf{Q}$ range, demonstrating that the coherent superlattice structures fully modulate the optical phonon band structures. Additionally, the optical phonon bands are split into four distinct sub-bands due to phonon wave interference over a superlattice period of four unit cells. Thus, the phononic energy band gaps can be estimated from the energy differences between adjacent sub-bands. **Figure. 3b** shows the adjacent longitudinal sub-bands measured in short-to-middle-wavelength regions ($q$ > 0.5), where the band gap reaches its maximum. Their energy separation yields a phononic band gap of 2.1 meV, in close agreement with the DFPT prediction of 1.98 meV.

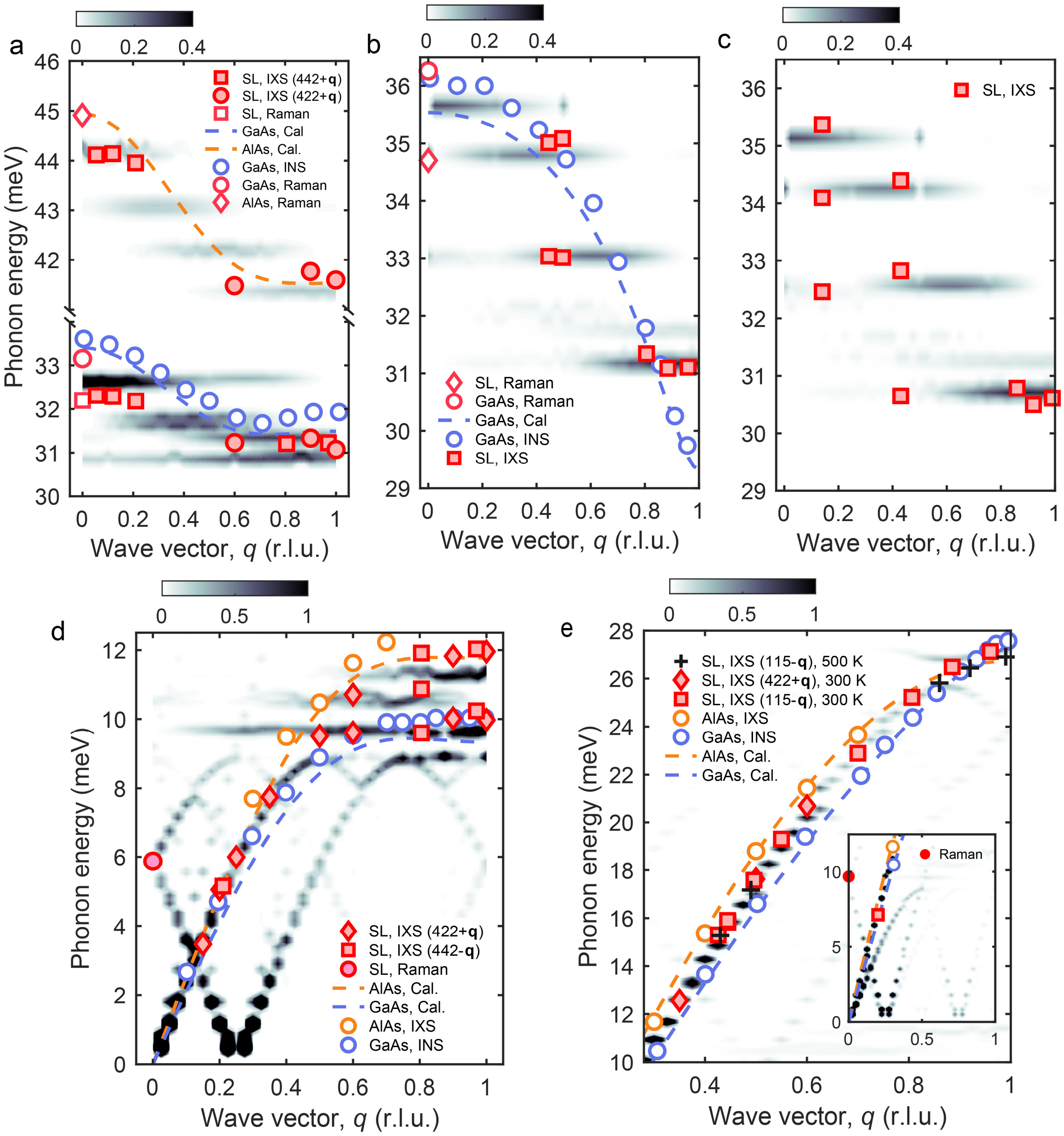

**Fig. 3. Coherently modulated phonon band structure and dynamic structure factor at 300 and 500 K along Γ-X direction**. (**a**) Coherently modulated transversal optical phonon bands of the superlattice (SL) at 300 K for **Q** = (4 4 2+*q*) and **Q** = (4 2 2+*q*). The measured coherently modulated phonon bands are consistent at the two BZs. (**b-c**) Coherently modulated longitudinal optical phonon bands for **Q** = (1 1 5-*q*) at 300 K (**b**) and 500 K (**c**). (**d**) Coherently modulated transversal acoustic phonon bands for **Q** = (4 4 2+*q*) and **Q** = (4 2 2+*q*) at 300 K. (**e**) Coherently modulated longitudinal acoustic phonon bands for **Q** = (1 1 5-*q*) at 300 and 500 K. The value of calculated $S(\mathbf{Q},\omega)$ in **Fig. 3a-e** is proportional to colormap values.

In addition, we performed Raman spectroscopy on a non-doped GaAs substrate, an AlAs thin film and our short-period superlattice at 300 K (**Fig. 3a-b** and **Fig. S3**). The results for GaAs and AlAs agree well with prior Raman[39] and inelastic neutron scattering measurements[40-43]. For the superlattice, the Γ-point optical phonon energies are lower than those of bulk GaAs, in agreement with our IXS results. This supports the phonon-band modulation associated with interference-induced phononic band gaps. In contrast to earlier studies[27,28] that reported multiple Raman peaks from folded phonon sub-bands, we observe only a single optical phonon mode at the Γ point. The Raman signals from the folded sub-bands are theoretically weak because they arise from higher-order scattering components[44]. Moreover, the optical sub-bands exhibit significant folding even with a short superlattice period of 2.26 nm (**Fig. 2a**). Strong three-phonon scattering at 300 K broadens these folded sub-bands, making their already-weak signals difficult to distinguish from one another.

To confirm the robustness of the coherently modulated phonon band structures, we increased the temperature to 500 K, at which the superlattice structure remained stable and phonon anharmonic scattering nearly doubled. As illustrated in **Fig. 3c**, the measured longitudinal optical phonon bands closely align with the $S(\mathbf{Q},\omega)$ calculated by DFPT, demonstrating that interference-induced phonon band modulation persists at 500 K. This robustness supports the potential application of coherent phonon band engineering at high temperatures.

Additionally, coherent modulation has been identified in the acoustic phonon bands. In particular, a 1.98-meV phononic band gap is clearly resolved in the transverse acoustic branch for $q > 0.9$, while a 1.1-meV phononic band gap is resolved at $q = 0.81$ and $q = 0.6$ (**Fig. 3d**). Moreover, the longitudinal acoustic branch of the superlattice lies between those of the two bulk materials at both 300 and 500 K (**Fig. 3e**), in excellent agreement with the modulated bands predicted by DFPT.

Raman spectroscopy provides further evidence of acoustic phonon band modulation in the superlattice. The spectrum (**Fig. S3e**) shows clear anti-Stokes and Stokes longitudinal acoustic phonon peaks at ±78 $cm^{-1}$ (±9.69 meV), corresponding to a zone-folded longitudinal acoustic sub-branch that originates at $q$=0.25 and folds back to the Γ point. The measured energy of 9.69 meV agrees well with the DFPT-calculated value of 9.3 meV (inset of **Fig. 3d**). Furthermore, a distinct Stokes peak at 47 $cm^{-1}$ (5.84 meV), corresponding to a zone-folded transverse acoustic branch, agrees well with the DFPT-calculated value of 5.75 meV (**Fig. 3e)**. These observations further support coherent modulation of the acoustic phonon bands in the superlattice.

Together, our IXS and Raman observations demonstrate that phonons in the superlattice follow the coherently modulated band structures rather than the dispersions of bulk materials. Thus, the superlattice no longer behaves as a composite of two materials, but as one homogeneous, fully coherent system suitable for phonon band engineering.

### Anharmonic phonon scattering in coherent superlattice structures

The coherently modulated phonon band structure of the superlattice profoundly impacts the three-phonon scattering lifetime ($\tau$), which is inversely proportional to the measured phonon linewidth ($\sigma=2\pi\hbar/\tau$). **Figure 4a** presents the mode-dependent three-phonon scattering phase space[45] ($SPS \propto \sigma$) for the three materials along the Γ-X direction. Compared with bulk GaAs and AlAs, the *SPS* of the superlattice nearly doubles in the energy ranges where its phonon bands flatten, particularly around 10 meV and above 29 meV. This enhancement suggests that band flattening increase the two-phonon density of states that account for energy and momentum conservations in three-phonon scattering. Consequently, the $\sigma$ of the superlattice is 1.5–2 times those of the two bulk materials, as shown by the *ab initio* lattice dynamics calculations and the IXS measurements in **Fig. 4b**.

Moreover, in coherent superlattice structures, $\sigma$ is modulated and scales linearly with $\omega$, in contrast to the typical $\omega^2$-dependence observed in bulk GaAs and AlAs. The $\sigma$ of acoustic phonons in the superlattice are therefore nearly one order of magnitude larger than that in bulk GaAs. The calculated cross-plane $\kappa$ of the superlattice (3.9 W/m·K at 300 K and 2.3 W/m·K at 500 K) is only 10% of that of bulk GaAs (46 W/m·K at 300 K and 21 W/m·K at 500 K). The cross-plane $\kappa$ measured by time-domain thermoreflectance (3.8 ± 0.3 W/m·K at 300 K and 2.4 ± 0.2 W/m·K at 500 K) agree well with the calculated values and validate the lattice dynamics approach. Because the superlattice band gaps do not reduce in-plane group velocities, the calculated in-plane $\kappa$ (7.9 W/m·K at 300 K and 4.8 W/m·K at 500 K) is less strongly suppressed than the cross-plane $\kappa$. Nevertheless, in-plane $\kappa$ remains well below that of bulk GaAs because the enhanced three-phonon scattering, occurring throughout the full three-dimensional reciprocal space, drastically reduces phonon lifetimes. These results demonstrate that modulated anharmonic phonon scattering significantly reduces both cross-plane and in-plane $\kappa$ in coherent superlattice structures above room temperature.

The phonon linewidth $\sigma$ and phonon energy shift $\Delta\omega$ can be derived from the imaginary and real parts of phonon self-energy[46], respectively. We decompose $\Delta\omega$ into $\Delta\omega_3 + \Delta\omega_4 + \Delta\omega_0$, where $\Delta\omega_3$ and $\Delta\omega_4$ are the phonon energy shift arising from cubic and quartic anharmonicity, respectively, and $\Delta\omega_0$ is the phonon energy shift arising from thermal expansion. The calculated and measured $\Delta\omega$ between 300 and 500 K are in reasonable agreement (**Fig. 4c**). Among these terms, $\Delta\omega_3$ is negatively correlated with $\sigma$ through the Kramers-Kronig relations[46]. It is typically negative and decreases with increasing phonon energy, making it a major contributor to the softening of the optical phonon bands despite $\Delta\omega_4$ being positive. Compared with bulk GaAs and AlAs, the enhanced *SPS* for optical phonons in the superlattice approximately doubles $\sigma$, potentially decreasing $\Delta\omega_3$ and strengthening the optical phonon softening effect in the superlattice. Thermal expansion also contributes substantially: the lattice parameter increases by a factor of 1.00159 from 300 to 500 K (**Fig. S4a-b**), resulting $\Delta\omega_0$ is comparable to $\Delta\omega_3$ (**Fig. 4c**).

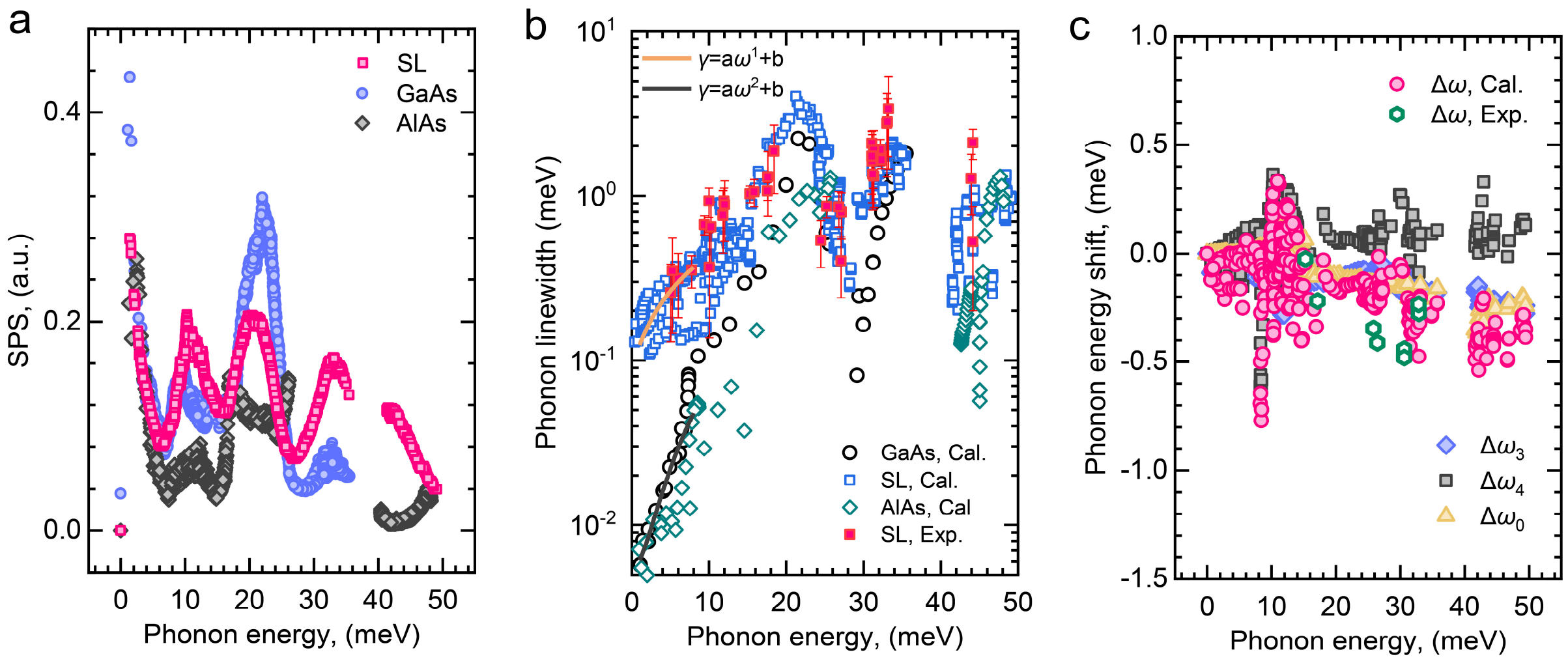

**Fig. 4. Anharmonic phonon scattering in coherent superlattice structures**. (**a**) Three-phonon scattering phase space of the superlattice (SL), GaAs, and AlAs. (**b**) Measured and calculated phonon linewidth ($\sigma$) of the superlattice in (0 0 1) direction, alongside the predicted $\sigma$ of GaAs and AlAs from *ab initio* lattice dynamics. (**c**) Phonon energy shifts from 300 to 500 K, with phonon energy at 300 K as the reference. In **Figs. 4b and 4c**, "Cal." and "Exp." denote the calculated and experimental results, respectively.

In brief, we have shown that that the coherently modulated phonon band structure increases *SPS* of optical phonons, further strengthening their three-phonon scattering rate and softening effects. We prospect this phenomenon to have significant implications for thermal transport, phononic metamaterials, electron-phonon coupling, and atomic-scale structural phase transitions[9,47].

## Conclusions

The interference of thermal phonon waves in artificial periodic nanostructures coherently modulates phonon band structures, laying the foundation for phonon band engineering. However, despite more than half a century of extensive research, direct observation of this phenomenon has remained elusive. Here, using high-resolution IXS, we directly observe coherently modulated thermal phonon band structure and lifetime in short-period GaAs/AlAs superlattices at 300 K. This modulation persists at 500 K, demonstrating robust coherent thermal phonon transport at high temperatures. Combined with *ab initio* lattice dynamics, our results reveal that thermal phonon interference not only modulates band structure but also expands the three-phonon scattering phase space, thereby strengthens anharmonic effects, such as three-phonon scattering and optical phonon softening. Our findings unveil new strategies for engineering phonon band structure and enhancing high-order phonon-phonon scattering through a flexible, bottom-up nanostructuring approach at high temperatures. This capability is crucial for optimizing phonon-mediated material properties and discovering phononic nanostructures with tailored phonon properties[48,49].

**Acknowledgments:** Y.L. thanks Terumasa Tadano for the fruitful discussions. This work was partially supported by CREST (Grant No. JPMJCR21O2) from Japan Science and Technology Agency (JST) and KAKENHI (Grant Nos. 22H04950, 21K14089 and 20H05660) from Japan Society for the Promotion of Science (JSPS). Y.L. thanks the Fellowship (Grant No. JP18J14024) from JSPS. Measurements at SPring-8 were performed under Proposals No. 2019B1666, No. 2020A0752, No. 2021A1487, No. 2022A2062, No. 2022A1318, and No. 2022B1407. The calculations in this work were partially performed using the supercomputer facilities at the Center for Computational Materials Science, Institute for Materials Research, Tohoku University.

**Author contributions:** J.S. directed the project; N.N. and K.H. grew the sample; N.M. and Y.A. performed wafer bonding; Y.L, H.U., H.F., T.M., R.N., D.I. and A.B performed inelastic X-ray scattering; Y.L., R.G., H.C., and B.X. performed Raman Spectroscopy and measured lattice thermal conductivity of the superlattice with time-domain thermoreflectance. Y.L. performed DFT and ab initio lattice dynamics calculations. Y.L. and J.S. wrote the manuscript. All authors commented on the manuscript.

**Competing interests:** Authors declare that they have no competing interests.

**Supplementary Materials**

Materials and Methods

Supplementary Text

Figs. S1 to S9